\documentclass[1p,authoryear]{elsarticle}
\usepackage{graphicx}
\usepackage{amsmath}
\usepackage{amssymb}
\usepackage{bm}
\newcommand{\vect}[1]{\boldsymbol{#1}}
\newcommand{\IMN}{Iwasa {\em et al}.~(2004)~}
\newcommand{\IMNe}{Iwasa {\em et al}.~(2004)}

\journal{Theoretical Population Biology}

\begin{document}

\begin{frontmatter}

\cortext[cor1]{Corresponding Author}

\title{The rate of mutli-step evolution in Moran and Wright-Fisher populations}

\author[ucsb]{Stephen R. Proulx \corref{cor1} }

\address[ucsb]{Ecology, Evolution and Marine Biology Department, UC Santa Barbara, Santa Barbara, CA 93106, USA}

\begin{abstract}
Several groups have recently modeled evolutionary transitions from an ancestral allele to a beneficial allele separated by one or more intervening mutants. The beneficial allele can become fixed if a succession of intermediate mutants are fixed or alternatively if successive mutants arise while the previous intermediate mutant is still segregating. This latter process has been termed stochastic tunneling. Previous work has focused on the Moran model of population genetics. I use elementary methods of analyzing stochastic processes to derive the probability of tunneling  in the limit of large population size for both Moran and Wright-Fisher populations. I also show how to efficiently obtain numerical results for finite populations.  These results show that the probability of stochastic tunneling is twice as large under the Wright-Fisher model as it is under the Moran model.

\end{abstract}

\begin{keyword}
Population Genetics \sep Stochastic Process \sep Stochastic Tunneling \sep Fixation Probability
\end{keyword}

\end{frontmatter}

\bibliographystyle{model2-names}

\section{Introduction}

Evolutionary biologists have long understood that transitions between adaptive sets of traits may involve multiple substitutions separated by neutral or maladaptive intermediate states \citep{wrightShift}. There has been a resurgence of interest in these ideas, in part because of advances in methods to measure epistatic interactions \citep[e.g. ][]{Tong1,Tong2} and  ability to observe evolutionary trajectories \citep{weinreich}. Several researchers have  modeled evolutionary processes when epistatic interactions allow for multiple genotypes to have the same direct effect on fitness but experience different evolutionary dynamics because of differences in their genetic robustness\citep{vanNimwegen,deVisser,ProulxCanalization,Draghi10} or the local mutational landscape \citep{EvoFlattest, OfallonAdlerProulx}. These scenarios can be called circum-neutral because alternative genotypes differ in their long-term evolutionary dynamics only because of the genomic circumstances in which they are found \citep{Flapping}. 

Several groups have extended the theory to describe the rates and probability of transition along a multi-step evolutionary pathway. \citet{weinreich} took the approach of calculating the total waiting time along various pathways and comparing the relative waiting times to reach a final state.  \citet{HermissonSoft} considered a scenario where previously accumulated genetic variation may become adaptive following an environmental shift. In this scenario the population genetic dynamics of standing variation plays an important role in determining how evolution proceeds at the next step in the process (see also \citet{KoppHermisson}).  \IMN derived approximate results on the waiting time and probability of a two-step sequence of mutational transitions using the Moran model, while  \citet{IMN03}  derived results in a Wright-Fisher model for a scenario where  multiple mutations are required to escape the immune response. These results  have been utilized by several other groups to study the rate of multi-step evolutionary processes \citep{DurrettSchmidt, LynchPNAS10,LynchAbegg}. Several other works have explored the probability and timing of multi-step processes, as well as exploring the validity of approximations \citep{Schweinsberg08,Weissman09,Durrett09}. Both \citet{Schweinsberg08} and \citet{Weissman09} have presented branching process approximations for large populations that are equivalent to the large population size limit results for the Moran model presented here.

The goal of this paper is to show how the finite population processes for both the Moran model and the Wright-Fisher model can be written and solved using  the method of first step analysis.
This  helps to clarify  some of the terms described by \IMN  and gives an algorithm for efficiently solving the finite population Moran model. 
The Moran tunneling probabilities have previously been applied to Wright-Fisher populations without verifying that these results still hold. I show that the Wright-Fisher tunneling probabilities differ from the Moran probabilities  by a factor of 2. This correction
will allow stochastic tunneling results to be applied to a wider range of scenarios. I also compare the large population size approximations for the rate of tunneling with simulations and exact calculations for small population size.

%Add bit on limitations of Iwasa approach

\subsection{Preliminary definitions and results}

By considering the population level evolutionary process as a series of transitions between populations fixed for a single genotype we can calculate the waiting time for the population to become fixed for secondary mutations. So long as $N \mu <<1$ we will seldom have multiple mutants arising in the same generation. This approach also assumes that each attempt at tunneling, if unsuccessful, is over before another primary mutant arises. Determining when this conditions actually holds is more difficult because the sojourn time of the primary mutant goes up as its selective disadvantage decreases. In the case of circum-neutral primary mutants, the sojourn times are characterized by large variances that become undefined as population size approaches infinity.  A rigorous analysis of the parameter combinations that allow this approximation to be applied is provided in \citet{Schweinsberg08} and \citet{Weissman09}. 

The first mutational step (the primary mutant) is assumed to have relative fitness $r\leq1$, while the second mutational step (secondary mutant) is assumed to have fitness $a>1$ relative to the ancestral allele. In the case where $r$ is exactly one, the first mutational step has no direct effect on fitness and the primary mutants can be considered circum-neutral \citep{Flapping}. Such circum-neutral substitutions do not directly affect reproductive fitness but do alter the long-term evolutionary trajectory of the population. The ancestral population can evolve to be fixed for the secondary mutant either through a sequential mutational pathway or because a lineage of primary mutants destined for extinction produces a secondary mutant which is destined for fixation, a process termed stochastic tunneling by \citep{Komarova03}.

The waiting time until a secondary mutation becomes fixed can be expressed in terms of the waiting times for the sequential and tunneling paths. I define the per generation probability of successful sequential substitutions $S_1$ and $S_2$ and the per generation probability of the opening of a successful tunnel as $T$. The waiting time for the transition between population states is well described by an exponential waiting time so long as population size is not too small \citep{Iwasa05}. This means that the process is characterized by a race between waiting for a primary mutation to arise and fix and the start of a tunneling pathway. The expectation of the total waiting time until a secondary mutation is given by
\begin{equation}
\label{eq:wait}
E[t] = \frac{T}{(T+S_1)^2} + \frac{S_1 (S_1+ S_2 +T)}{S_2 (T + S_1)^2},
\end{equation}
where the first term represents the contribution to the expected waiting time from tunneling pathways and the second term represents the contribution from sequential pathways. If $T=0$ this is simply the sum of the waiting times for primary and secondary mutations to sequentially fix. This approximation ignores the time that it takes for beneficial mutations to spread through the population and the amount of time that primary mutants are segregating before a secondary mutation arises. The time required for alleles destined to fix to spread to fixation is typically much smaller than the waiting times for them to arise, and in any case it can be simply added to the total waiting time \citep[see][]{LynchAbegg}.

The per generation probabilities of sequential fixation are 
\begin{eqnarray}
S_1 &=& N \mu_1 U(r) , \\
S_2 &=& N \mu_2 U(a/r),
\end{eqnarray}
where $N$ is the haploid population sizeÊ(for simplicity I assume this is approximately the effective population size as well), $\mu_1$ is the probability that an ancestral allele will mutate into a  primary mutant, $\mu_2$ is the probability that the primary mutant will mutate into a secondary mutant, and $U(x)$ is the fixation probability of a mutation with relative fitness $x$ when initially present as a single copy. Because this follows sequential fixation of mutants, the secondary mutant  is invading into a population fixed for the primary mutant, giving it a relative fitness of $a/r$.

Following \IMNe, the probability of tunneling can be written as 
\begin{equation}
T = (1-U(r)) (1-E[\mathrm{no\ secondary\ substitution} | \mathrm{extinction}]),
\end{equation}
where $E[\mathrm{no\ secondary\ substitution} | \mathrm{extinction}]$ represents the probability that no successful secondary mutations arise while the primary mutant is segregating conditioned on the eventual extinction of the lineage of primary mutants. This can be related to the unconditional expectation  by 
\begin{equation}
\label{eq:CondT}
E[\mathrm{no\ secondary\ substitution}  | \mathrm{extinction}] = E[\mathrm{no\ secondary\ substitution} ] (1-U(r))
\end{equation}
 \citep{Iwasa04}. This provides a simple relationship between  calculations made  using the conditioned trajectory of primary mutations and the unconditioned trajectory of primary mutations.

% Equation \ref{eq:wait} is a

\section{Moran Model}
%This is where the generic finite pop size equations go
The Moran model \citep{Moran} follows a population of size $N$ described by a vector $\vect{x}$ where $x_i$ indicates the number of individuals of genotype $i$. In the Moran process, each unit of time either 2 elements of $\vect{x}$ change by one unit each in opposite directions or $\vect{x}$  remains constant. This model is often conceptually described as one where an individual is chosen to reproduce at random  weighted by their reproductive output. Population size is kept constant by choosing one of the original population members to die (it may be the one that reproduced). Mutation causes offspring to differ from their parent's genotype with a probability defined by the mutation rate. The scale of time in this model is population size dependent; a generation is measured in terms of $N$ time steps.

Because the number of individuals of each genotype can change by at most one unit each time step, the Moran model can be expressed as a Markov chain whose matrix definition is tridiagonal. This is in sharp contrast to the Wright-Fisher model where any population state can move to any other state in one generation (albeit with low probabilities). This simple matrix structure allows many features of the stochastic process to be expressed algebraically.

The evolution of the population can be described as a series of transitions between states described by the complement of segregating alleles. Assuming that primary mutations occur rarely enough so that multiple primary mutants do not typically arise together, the transition probabilities between states can be based on the introduction of a single individual mutant. As this approximation breaks down more error will be introduced into the transition probabilities. 
The ancestral population is monomorphic for the ancestral genotype. Following the introduction of a primary mutant, the population will evolve with two genotypes for some time until either the lineage of primary mutants goes extinct or a secondary mutant lineage arises and does not go extinct. 

Following the introduction of a primary mutant, the population is composed of $i$ primary mutants and $N-i$ ancestral alleles. In the absence of mutation, the Markov transition probabilities are 
\begin{eqnarray*}
Pr(i\rightarrow i-1) &=& \frac{i (N-i)}{N (i r + (N-i))}\\
Pr(i\rightarrow i) &=& \frac{   i^2 r + (N-i)^2       }{N (i r + (N-i))}\\
Pr(i\rightarrow i+1) &=& \frac{i r (N-i)}{N (i r + (N-i))},
\end{eqnarray*}
where $0<i<N$ and $r$ is the relative fitness of the primary mutant. Note that this model ignores the change in the number of primary mutants due to their mutation into secondary mutants.  Following the introduction of the primary mutant, it may produce a lineage that eventually goes extinct or eventually becomes fixed in the population. It is useful to describe the population process conditioned on the eventual extinction of the primary mutation in order to consider these two scenarios separately. The conditional process can simply be described by
\begin{equation*}
\Pr(i\rightarrow j |\mathrm{extinction}) =\Pr(i\rightarrow j) \frac{\pi_j}{\pi_i} 
\end{equation*}
where $\pi_i$ represents the the probability that the lineage goes extinct given that there are currently $i$ mutants in the population \citep{Ewens73}.

I use first step analysis \citep{TaylorKarlin} to in order to find the total probability that no successful secondary mutant is spawned from a lineage of primary mutants destined to eventual extinction. Let $v_i$ be the probability that no successful secondary mutants are spawned from a lineage beginning with $i$ mutants.  $v_i$ can be implicitly defined as the probability that no successful mutants arise in the current time step multiplied by the probability that no successful mutants are produced in the future. For the process conditioned on eventual extinction of the primary mutant lineage we have 
\begin{eqnarray}
\label{eq:Sys}
v_i = (1- \omega r \frac{i}{N})\left( Pr(i\rightarrow i-1) \frac{\pi_{i-1}}{\pi_i} v_{i-1} + Pr(i\rightarrow i) \frac{\pi_{i}}{\pi_i} v_{i} + Pr(i\rightarrow i+1) \frac{\pi_{i+1}}{\pi_i} v_{i+1} \right), 
\end{eqnarray}
where the composite parameter $\omega = \mu_2 U(a)$. Note that $v_0=1$ and $\pi_{N}=0$. This then is a system of $N-1$ linear equations with $N-1$ unknowns. The probability of tunneling is simply $T=N \mu_1 (1-v_1)  \pi_1$.

\subsection{Algorithm for solving the finite population size model}
For finite populations the system of equations can be represented as a tridiagonal matrix. This system can be solved numerically using a  mathematical computing package. Many results are known for the matrix inverse and eigenvectors of tridiagonal matrices \citep{Usmani, daFonseca}. These results can be used to numerically calculate the eigenvectors for even large population size because they involve less than $4 N$ multiplications and additions. Because they use recursive calculations there is no constraint imposed by memory levels and computation time basically increases linearly. On a MacPro with a 3.2 GHz Xeon processor running Mathematica 7 the calculation for population size of $10^6$ takes about 50 seconds.

Given our system of equations (\ref{eq:Sys}) we can write a matrix equation of the form
\begin{equation*}
\mathbf{A} \mathbf{v} = \mathbf{x},
\end{equation*}
where $\mathbf{v}$ is the vector of $v_i$ and 
\begin{align*}
\mathbf{A}& = \begin{pmatrix}  a_1 & b_1 \\
                                                    c_1 & a_2 & b_2\\
                                                            & c_2 & a_3 & b_3\\
                                                            &         & c_3 & \ddots & \ddots\\
                                                            &         &         & \ddots & \ddots & b_{N-2}\\
                                                            &         &         &             & c_{N-2} & a_{N-1}
                                                    \end{pmatrix} \\
\mathbf{x} &=                                                     \begin{pmatrix} x_1 \\ 0 \\ \vdots \\ 0 \end{pmatrix} .
\end{align*}
The elements of the matrix can be found from equation (\ref{eq:Sys}) and I provide formulas for $a_i$, $b_i$, $c_i$ and $x_1$ below.
Note that this system has $N-1$ rows because it is conditioned on the eventual extinction of the primary mutant.  
 \citep{daFonseca} defines the recursive equations
 \begin{align*}
 \theta_i &= a_i \theta_{i-1} - b_{i-1} c_{i-1} \theta_{i-2}, \\
 \theta_0 &= 1, \\
 \theta_1 &=a_1 \\
 \phi_i &= a_i \phi_{i+1} - b_i c_i \phi_{i+2} ,\\
 \phi_N &= 1 ,\\
 \phi_{N-1} &= a_{N-1} .
 \end{align*}
Each element of $\mathbf{A}^{-1}$ can be expressed as an algebraic expression of $\theta$ and $\phi$.  Because $v_1$ is the first element of $\mathbf{A}^{-1} \mathbf{x}$ and because only the first element of $\mathbf{x}$ is non-zero we need only calculate $\mathbf{A}^{-1}_{1,1}$. 
\begin{equation}
\mathbf{A}^{-1}_{1,1} = \frac{\theta_0 \phi_{2} }{\theta_{N-1}} . 
\end{equation}

Given the system of equations (\ref{eq:Sys}) we have
\begin{align*}
x_1 &= \frac{(N-1)(N-\omega r)}{N^2(N-1+r)(1-\pi_1)} \\
a_i &= \frac{(N-i)^2 +i^2 r}{N(N-i+i r)} \frac{N- i r \omega}{N} -1 \\
b_i &= \frac{(N-i) i r (1-\pi_{i+1})}{N(N-i+i r)(1-\pi_{i})} \frac{N- i r \omega}{N}  \\
c_i &= \frac{(N-i-1)(i+1)(1-\pi_{i})}{N(N-i + i r)(1-\pi_{i+1})} \frac{N- (i+1) r \omega}{N}
\end{align*}
Finally, the total probability that no successful secondary mutations are produced while a lineage descending from a single primary mutant is extant is
\begin{equation}
v_1 = \frac{ \phi_{2} }{\theta_{N-1}}  \left(- \left(1-\frac{r \omega}{N}\right)  \frac{N-1}{N(N-1+r)(1-\pi_1)} \right). 
\end{equation}
This method can be used to numerically solve for $v_1$ and is reasonably quick even in large populations.

\section{Wright-Fisher Populations}

\subsection{Finite Population Size}
In the Wright-Fisher formulation the population at generation $t+1$ is found by sampling gametes produced in generation $t$. So long as the number of gametes produced is reasonably large, the probability distribution for adults in generation $t+1$ is binomial such that
\begin{align*}
Pr(0 \rightarrow 0) &= 1 ,\\
Pr(i\rightarrow j) &= \binom{N}{j} \left(\frac{i r}{i r +(N-i)}\right)^j \left(1-\frac{i r}{i r +(N-i)}\right)^{N-j} , \\
Pr(N \rightarrow N) &= 1 ,\\
\end{align*}
represents the probability that the population goes from $i$ mutants to $j$ mutants in one generation. Again I use a first step analysis to calculate the probability that no secondary mutations arise beginning from a single mutant. Exact calculations of the fixation probabilities for the finite Wright-Fisher model are not available, so this system cannot be converted to the process conditioned on eventual extinction of the primary mutant. This means that the probability of tunneling will have to be back-calculated  from equation (\ref{eq:CondT}). If the primary mutant becomes fixed then the probability that a successful secondary mutation is spawned is 1. 
 
I define $\tilde{v}_i$ as the probability that no successful secondary mutations are spawned  starting from the state where $i$ primary mutants are present. For each possible state $i$, the probability that no successful secondary mutants are produced is the probability that none of the $i$ primary mutants immediately produce a successful secondary mutant ($(1-\omega)^i$) multiplied by the sum of the probabilities that the next generation contains $j$ primary mutants multiplied by the probability that a lineage starting with $j$ mutants never produces a successful secondary mutant.  This is slightly different from the Moran model where only one secondary mutant can possibly arise at each time point.
This gives
\begin{align*}
\tilde{v}_0 &= 1 ,\\
\tilde{v}_i &=(1 -\omega)^i \sum_{j=0}^{N} Pr(i\rightarrow j) \tilde{v}_j , \\
\tilde{v}_N &= 0.
\end{align*}
The equations for $1\leq i \leq N-1$  can be rewritten as a sum of $\tilde{v}_j$ terms as follows
\begin{align*}
0 &= (1 -\omega)^i \sum_{j=0}^{N} Pr(i\rightarrow j) \tilde{v}_j -\tilde{v}_i\\
    & =(1 -\omega)^i \left (\sum_{j=0}^{i-1} Pr(i\rightarrow j) \tilde{v}_j + Pr(i\rightarrow i) (\tilde{v}_i -1) +\sum_{j=i+1}^{N} Pr(i\rightarrow j) \tilde{v}_j \right).
\end{align*}
This is a system of linear equations in $\tilde{v}_j$ and can be  written in matrix form as $\mathbf{A} \mathbf{\tilde{v}} = \mathbf{x}$ where
\begin{align*}
\mathbf{A}& = \begin{pmatrix}  1  & 0 &\hdots & \hdots &0 \\
       (1 -\omega)^1  Pr(1 \rightarrow 0)   & (1 -\omega)^1  Pr(1 \rightarrow 1) -1   & (1 -\omega)^1  Pr(1 \rightarrow 2) & \hdots & \hdots \\
        (1 -\omega)^2  Pr(2 \rightarrow 0)   & (1 -\omega)^2  Pr(2 \rightarrow 1)    & (1 -\omega)^2  Pr(2 \rightarrow 2) -1 &\hdots & \hdots\\
        \vdots & \vdots & \vdots &\ddots&\vdots \\
%                      \vdots                                      &    \vdots     & \vdots   & \ddots &  (1 -\omega)^{N-1}  Pr(N-1 \rightarrow N-1) -1 &  (1 -\omega)^{N-1}  Pr(N-1 \rightarrow N)
                      0                                      &    0     & \hdots   & \hdots &   1
                                                                          \end{pmatrix} 
\end{align*}
\begin{align*}
 \mathbf{\tilde{v}} =                                                     \begin{pmatrix} \tilde{v_1} \\ \tilde{v_2} \\ \vdots \\ \tilde{v_N} \end{pmatrix} & & \mathbf{x} =                                                     \begin{pmatrix} 1 \\ 0 \\ \vdots \\ 0 \end{pmatrix}.
\end{align*}
The solution to  $\mathbf{\tilde{v}} = \mathbf{A}^{-1} \mathbf{x}$  can be found numerically using standard mathematical packages. Compared to the solutions for the Moran model, the Wright-Fisher model requires many more computational operations for the same population size.

Recall that $\tilde{v}_1$  is the unconditioned probability that no successful secondary mutations are produced from a lineage of primary mutants descending from a single primary mutant. We can calculate the probability that no successful secondary mutations are produced conditioned on the eventual extinction of a lineage of primary mutants descending from a single initial mutant using the approximate fixation probability for a single copy mutant \citep{CrowKimura} and  equation (\ref{eq:CondT}) as
\begin{equation*}
v_1 = \tilde{v}_1 \left(1-\frac{1-e^{-2(r-1)}}{1-e^{-2 N (r-1)}} \right).
\end{equation*}

%There is an example of the finite pop binomial model in the file "FredSolutionPoisson"
\section{Large Population Size Approximations}
In very large populations, the dynamics of newly introduced mutants can be modeled as a branching process. In this limit, segregating mutants do not interact and mean fitness is not altered by their spread in the population. The probability that a newly introduced primary mutant leads to the eventual maintenance of a secondary mutant can then be calculated based on this branching process, while fixation probabilities of beneficial alleles may still be modeled using finite population size results. In other words, population size enters the calculations in two ways; the finite population size causes frequency dependent interactions between segregating mutant alleles and causes the fixation probability of beneficial mutants to increase. The large population size approximation takes this first population size to be large while leaving this second population size finite. 

The main feature of the branching process approximation that allows this to be calculated is that the probability that no secondary mutations persist as a function of the number of initial mutants scales as $v_i = \alpha^i$, where $\alpha$ represents the probability that no secondary mutations persist when a single primary mutant is introduced. Once $\alpha$ has been solved for the per generation probability of tunneling is described by
\begin{equation}
T= N \mu_1 (1-\alpha) (1-U(r)) .
\end{equation}

\subsection{Moran Model}
The calculation of the expected probability of a successful secondary mutation in the  large population limit begins with the same first step analysis equations as in a finite population. Starting with equation (\ref{eq:Sys}) setting $i=1$ and taking the limit as $N\rightarrow \infty$ we have 
\begin{eqnarray}
\label{eq:v2M}
v_2 &=& \frac{v_1 (1+r(1+\omega))-1}{r}. 
\end{eqnarray}
Inserting $v_i = \alpha_{\mathrm{M}}^i$ into equation (\ref{eq:v2M}) gives
\begin{eqnarray}
\label{eq:alphaM}
\alpha_{\mathrm{M}} &= \frac{1+r(\omega+1) - \sqrt{(1-r)^2 + (2+r(\omega+2)) r \omega } }{2 r} .
\end{eqnarray}

Note that this is slightly different from the formula  Iwasa {\em et al.} (2004) give because theirs is an expression in terms of the unconditional process. Recalling that $E[P | \mathrm{extinction}] = E[P] (1-U(r))$ their expression can be rewritten in terms of the conditional process as
\begin{equation}
\alpha_{\mathrm{IMN}} = \frac{2-\sqrt{(1-r)^2+2(1+r) r \omega}}{1+r}(1-U(r)).
\end{equation}
In general, these equations are extremely similar and will often be so close as to be numerically indistinguishable for any empirical applications. Notable exceptions are when the population is small enough that the \citet{Iwasa04} breaks down and $\alpha_{\mathrm{IMN}}>1$.

In the case of a circum-neutral primary mutation ($r=1$) the two expressions become
\begin{eqnarray}
\label{eq:MoranAlpha}
\alpha_{\mathrm{M}} &=& 1+\frac{1}{2} \left( \omega-\sqrt{\omega(4+\omega)}\right) \\
\alpha_{\mathrm{IMN}} &=&\left(1-\sqrt{\omega}\right)\frac{N-1}{N}.
\end{eqnarray}
Recalling that $\omega = \mu_2 \frac{1-1/a}{1-1/a^N} $ and taking the limit as $N\to \infty$ we get
\begin{eqnarray}
&\lim_{N\to\infty} \alpha_{\mathrm{M}}  &=  1-\sqrt{(1+\frac{a-1}{2 a} \mu_2)^2-1}+ \frac{a-1}{2 a} \mu_2 \\
&\lim_{N\to\infty} \alpha_\mathrm{IMN} &= 1-\sqrt{\frac{a-1}{a} \mu_2}
\end{eqnarray}
(See \citep{Schweinsberg08} for an alternate derivation). These expressions are most different when $a$ is large and $\mu_2$ is large. However, even for an unrealistically large value $\mu_2=10^{-2}$ the expressions are different by less than $1\%$.

\subsection{Wright-Fisher Model}

\begin{figure}[htbp]
\begin{center}
%\begin{tabular}{cc}
\includegraphics[width=0.8\textwidth]{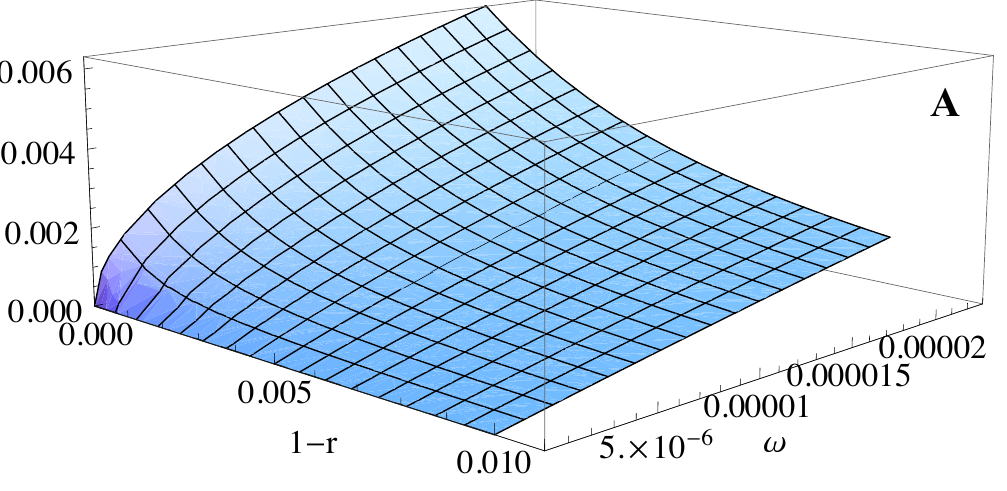} 
\includegraphics[width=0.8\textwidth]{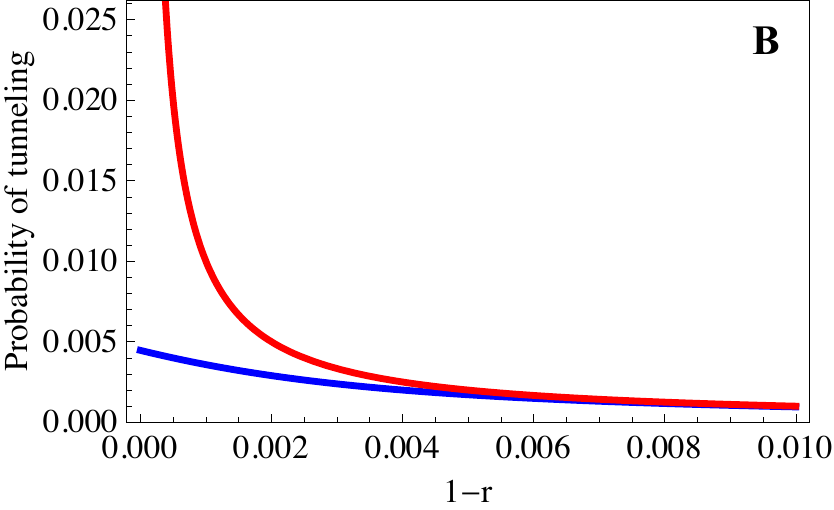}
%\end{tabular}
\caption{The solution for the probability of tunneling in large Wright-Fisher populations.  Panel A shows the probability of tunneling as a function of $\omega$ and $1-r$ (note that  the selection coefficient $s=1-r$). When $1 - r\approx 0$ the curve is approximately given by equation (\ref{eq:NeutWF}). As $1-r$ increases the probability of tunneling approaches that given by equation (\ref{eq:BalanceWF}). Panel B shows the decrease of the probability of tunneling as $1-r$ goes up in blue ($\omega=0.00001$). The red curve shows the approximate value for small $\sqrt{\omega}$ relative to $1-r$ from equation (\ref{eq:BalanceWF}). For this value of $\omega$, the two curves are within $5\%$ once $1-r > 0.01$. 
%for second panel omega is fixed at 0.00001. It has an accuracy of 0.05 at around s=0.01
}
\label{fig:SimWF3D}
\end{center}
\end{figure}

Using the branching process approximation we have $v_j = \alpha_{\mathrm{WF}}^j $. For $1\leq i\leq N-1$ this gives
\begin{align}
\label{eq:vbin}
v_i &=(1 -\omega)^i \sum_{j=0}^{N} \binom{N}{j} \left(\frac{i r}{i r +(N-i)}\right)^j \left(1-\frac{i r}{i r +(N-i)}\right)^{N-j} \alpha_{\mathrm{WF}}^j .
\end{align}
For any specific $i$, $v_i$ can be approximated by noting that the binomial probabilities approach a Poisson distribution as $N\rightarrow \infty$ \citep{RossBook,IMN03}. Rewriting equation (\ref{eq:vbin}) in the limit of large $N$ using the Poisson approximation gives
\begin{align}
\label{eq:vpois}
v_i & = (1 -\omega)^i \sum_{j=0}^{\infty} \frac{e^{-i r } (i r)^j}{j!} \alpha_{\mathrm{WF}}^j &= (1 -\omega)^i \left(e^{- r (1-\alpha_{\mathrm{WF}})} \right)^i.
\end{align}
A similar result was obtained by \citet{IMN03} for multi-step pathways involving multiple routes to a higher fitness mutant.
For a primary mutant lineage starting with 1 individual the probability of tunneling is the solution of 
\begin{align}
\label{eq:ImpWF}
\alpha_{\mathrm{WF}} &= e^{- r (1-\alpha_{\mathrm{WF}})} (1-\omega).
\end{align}
This implicit definition of $\alpha_{\mathrm{WF}}$ can be easily solved numerically for specific values of $r$ and $\omega$ (figure \ref{fig:SimWF3D}).

The implicit equation for the probability that no successful secondary mutants are produced has a convenient interpretation. In large Wright-Fisher populations the distribution of offspring produced by a single parent is Poisson. A newly arising primary mutant has probability $\omega$ of immediately mutating and spawning a lineage of  secondary mutants that is destined to fix. If it does not (with probability $1-\omega$) then it produces a Poisson distributed number of offspring mutants with mean $r$, each of which has an effectively independent probability $1-\alpha_{\mathrm{WF}}$ of spawning a secondary mutant lineage itself destined to fix. The probability that none of these primary mutant offspring spawn a successful secondary mutant lineage is simply the 0 term from the Poisson distribution with parameter $r(1-\alpha_{\mathrm{WF}})$. Thus, $\alpha_{\mathrm{WF}}$ is equal to the product of the probability that the lone primary mutant does not immediately produce a successful secondary mutant with the probability that none of the primary mutant offspring produce a successful secondary mutant lineage. 

%another 3D plot of tunneling probability. Probably a panel w/the other 3D plot.

 \subsection{Comparisons between the models}
 
%These Taylor's series approximations are in IncludingSelection.nb

Define $T_{\mathrm{M}}$ and  $T_{\mathrm{WF}}$ as the probability that a successful secondary mutation arises from a lineage of primary mutants founded by a single primary mutant in the Moran and Wright-Fisher models, respectively. 

For the Moran model where $r=1$,  $T_{\mathrm{M}} = \sqrt{\omega(1+\omega/4)}-\frac{\omega}{2}$ (i.e. $1-\alpha_{\mathrm{M}}$ from equation (\ref{eq:MoranAlpha})). For small $\omega$ this can be approximated using a series expansion (see appendix \ref{sec:series}). Likewise the Wright-Fisher model can be approximated using equation (\ref{eq:ImpWF}) and again expanding around small $\omega$ (see appendix \ref{sec:series}).
\begin{align}
\label{eq:NeutM}
T_{\mathrm{M}} & \approx  \sqrt{\omega} -         \frac{\omega}{2}  + O[\omega]^{3/2} , \\
\label{eq:NeutWF}
T_{\mathrm{WF}} &\approx \sqrt{2 \omega} - \frac{2 \omega}{3} + O[\omega]^{3/2} .
\end{align}
Noting that the fixation probability approaches $U(a) = (a-1)$ in large Moran populations and $U(a) = 2(a-1)$ in large Wright-Fisher populations and substituting $\omega=2 \mu_2 U(a) $ gives 
\begin{align}
T_{\mathrm{M}} &\approx  \sqrt{\mu_2 (a-1)} -         \frac{\mu_2(a-1)}{2}   ,\\
T_{\mathrm{WF}} &\approx 2 \sqrt{ \mu_2 (a-1) } - \frac{4 \mu_2 (a-1)}{3} .
\end{align}
These calculations show that for the same level of $\mu_2$ and $a$ the Wright-Fisher tunneling probability is a factor of 2 larger than for the Moran model.

\begin{figure}[htbp]
\begin{center}
\includegraphics[width=1.0\textwidth]{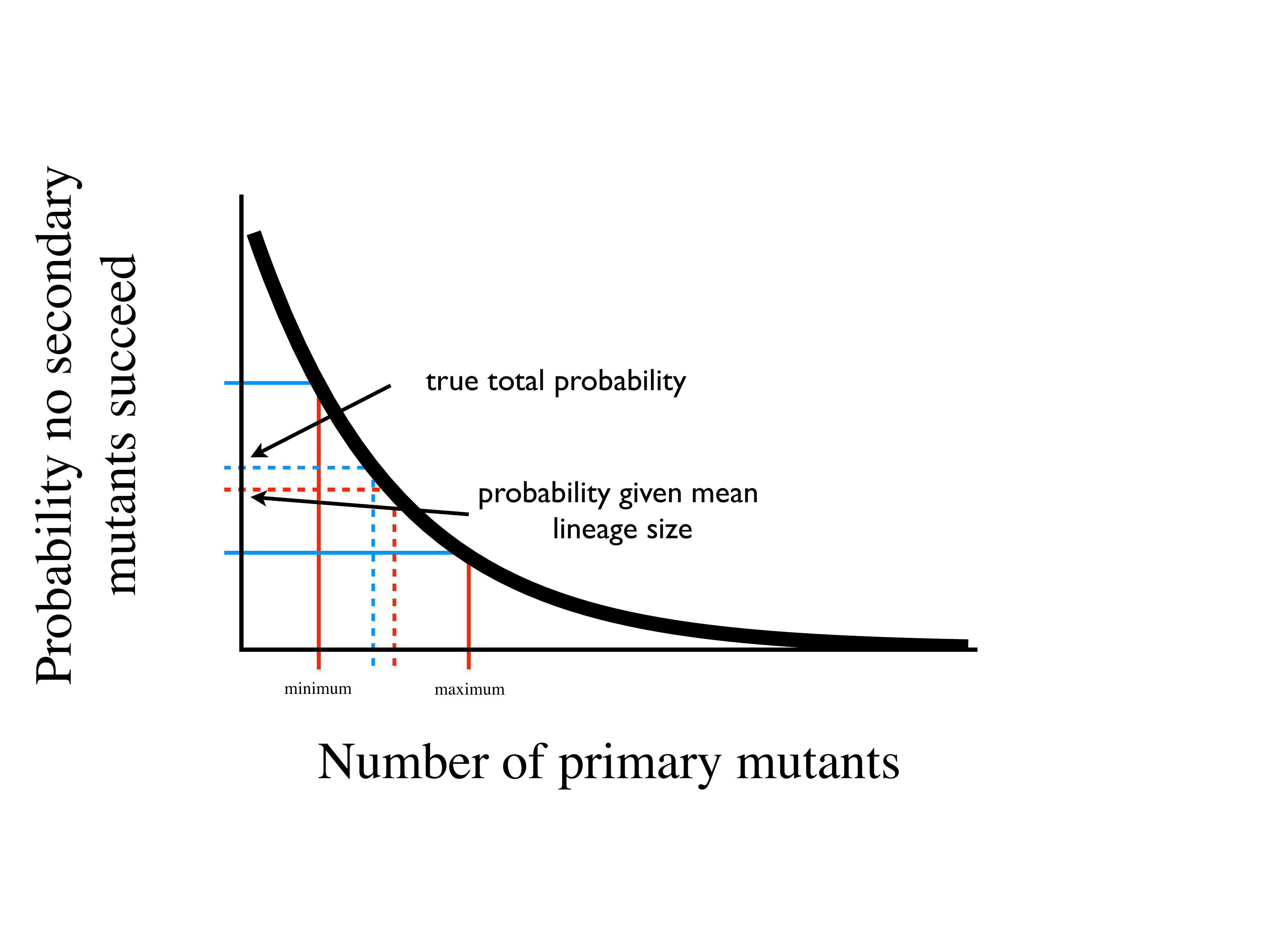}
\caption{A schematic illustration of the relationship between the probability of tunneling and the mean number of primary mutants. The probability that no successful secondary mutants are produced is a decreasing function of the primary mutant lineage size with positive second derivative. Assume that the primary mutant lineage takes on two possible values with equal probability (the minimum and maximum). The red lines map from the primary mutant lineage size and have a mean given by the red dashed line. The probability that no successful secondary mutant is produced from events having the mean lineage size is shown where the red dashed line meets the vertical axis. The true total probability that no successful secondary mutants are produced is found by averaging the probabilities  in the two different lineage sizes which can be visualized by finding the midpoint between the blue horizontal lines. Thus, the true probability that no secondary mutants are produced is higher than that found from the mean population size. This means that the probability of tunneling is smaller than would be expected based on mean population size alone.}
\label{fig:jensen}
\end{center}
\end{figure}

In both models the probability of tunneling depends on the distribution of the number of primary mutants spawned by a lineage founded by one primary mutant and the curvature of the function relating the probability a successful secondary mutant arises to the total number of primary mutants in a given lineage. In the limit of large population size, where a branching process approximation is valid, this distribution can be exactly determined \citep{Dwass}. Examining the relationship between the expectation of the probability no successful secondary mutant arises and the distribution of the primary mutant lineage size can help to understand this effect. Once population size gets to be reasonably large, the distribution of primary mutant lineage size remains almost constant except at the tail. In the Wright-Fisher case, when $r=1$ we have a critical branching process. Conditioned on eventual extinction of the primary mutant lineage, we can calculate the expected number of primary mutants spawned in a lineage arising from a single primary mutant  (see  \ref{sec:MeanSpawned}). Alternatively, we could calculate the distribution of primary mutant lineage sizes and then calculate the total probability that no successful secondary mutations are spawned. Figure \ref{fig:jensen} shows how the curvature of the function for the probability of successful secondary mutations as a function of primary mutation lineage size would lead to a reduced total probability that a secondary mutation becomes fixed. Because of Jensen's inequality, the expectation of the function of the random variable is larger than the function of the expectation. Because the probability cannot be below 0, no matter how large the primary mutant lineage, the contribution of very large primary mutation lineages is negligible. Thus, $T$ is considerably smaller than we would predict based only on the expected number of primary mutants spawned by a lineage destined to eventually become extinct.

For cases where $r<1$ approximating around small $\omega$ is more straightforward  (see appendix \ref{sec:series}). Approximating and substituting in the values for $\omega$ gives
\begin{align}
\label{eq:BalanceM}
T_{\mathrm{M}} &\approx   \mu (a-1)  \frac{r}{1-r}   \\
\label{eq:BalanceWF}
T_{\mathrm{WF}} &\approx 2 \mu (a-1)  \frac{1}{1-r}  .
\end{align}
For $r$ near one this approximation fails breaks down and either the exact equation or the neutral approximation should be used (figure \ref{fig:SimWF3D}).
Again, the probability of tunneling is larger for Wright-Fisher populations, but now they are different by a factor of $2 r$. The factor of 2 is again because of higher fixation probabilities in the Wright Fisher model while the factor of $r$ is due to the Moran model assumption that mutations only occur during reproduction. If, alternatively,  mutations were assumed to happen at a constant rate per generation (where generations encompass $N$ elementary steps of the Moran process), then the results would only differ because of their fixation probabilities (to the first order approximation).  

These results agree with the Moran population results of \IMNe. \IMN characterize the regime where $r<1$ by arguing that it arises as the product of the equilibrium frequency of the deleterious primary mutation, the probability a secondary mutant arises, and the probability of successful fixation of the secondary mutation. The results presented here provide a different interpretation. The probability of tunneling as derived here is the average over independent trajectories following the introduction of a single mutant. Such trajectories are never at equilibrium, but do spawn a characteristic distribution of mutants before their lineage goes extinct. In both Moran and Wright-Fisher populations, deleterious mutants produce an average of $\frac{1}{1-r}$ descendants (see \ref{sec:MeanSpawned}).  An important caveat is that equations (\ref{eq:BalanceM}) and (\ref{eq:BalanceWF}) only apply  when $r<1$ and $\omega$ is small. In these cases equations (\ref{eq:alphaM}) and (\ref{eq:ImpWF}) should be used instead.

\section{Multiple intermediate steps}
This approach readily extends to the case where multiple intermediate mutational steps stand between the ancestral state and any mutants that have improved fitness. Such multi-step probabilities have been calculated for the Moran model by \IMN, \citet{Schweinsberg08},  \citet{LynchAbegg,LynchPNAS10}, and \citet{Weissman09}.  \cite{Schweinsberg08} considered the neutral case and found that the approximation is valid when $\mu$ is small relative to the inverse of the population size squared. \citet{LynchPNAS10} considered both the multistep tunneling probability for both neutral and deleterious intermediates but did so by assuming that each intermediate deleterious mutation first rose to mutation selection balance. Additionally, \citet{Iwasa04} considered a immune response escape mutants in a multi-step Wright-Fisher model. 
\citet{Weissman09} examined the probability of multi-step tunneling for an arbitrary number of intermediate mutations and rigorously defined the regions of parameter space where the stochastic tunneling approximation is valid. They note that the stochastic tunneling approximation, as used here, is only valid when population size is large. Specifically, the required population size goes up if there are many intermediate mutations and they are only weakly selected against \citep{Weissman09}.  

The probability that a multistep tunnel is opened can be calculated recursively.  Using equation (\ref{eq:alphaM}), but writing  $\omega$ generically as  $\mu U$, I define a function 
\begin{equation}
\label{eq:MRecur}
g(\mu, U) = 1- \frac{1+r(\mu U+1) - \sqrt{(1-r)^2 + (2+r(\mu U+2)) r \mu U } }{2 r}. 
 \end{equation}
For a 2-step process, the probability of a tunnel is simply $g(\mu_2, U(a))$. Further recursions give the total probability for longer tunnels, so that a three-step tunnel opens with probability $g(\mu_2,g(\mu_3,U(a)))$. 

\begin{figure}
\begin{center}
\includegraphics[width=0.8\textwidth]{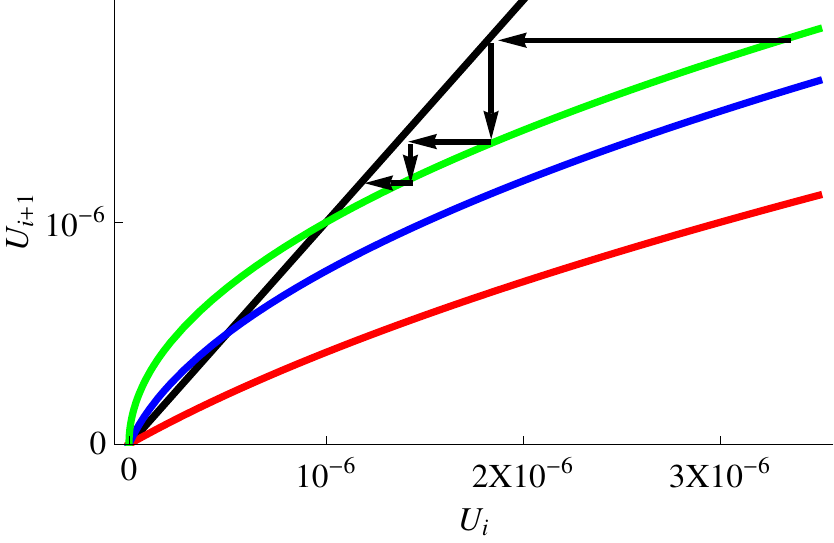}
\caption{The probability of tunneling in multistep pathways under the Moran model can be found by recursively applying the formula for the probability of tunneling. The diagonal line is shown in black along with the recursion formula from equation (\ref{eq:MRecur}) for three different values of $r$. In all cases, $\mu=10^{-6}$. The green curve is for $r=1$, where each intermediate mutation has the same fitness as the ancestral allele. The graph can be cobwebbed as shown by the black arrows. The initial value is the probability that the beneficial mutant at the end of the pathway becomes fixed, starting from a single individual. Regardless of the initial condition, the probability of tunneling through a long pathway converges on the value where the green curve crosses the black line. The blue curve is for slightly deleterious intermediate mutants with $r=0.9999995$. This still crosses the black line at a positive value. The red curve is for a lower value of $r=0.999998$. The red curve is always below the black line, so longer pathways have essentially no chance of fixing via tunneling. }
\label{fig:cobweb}
\end{center}
\end{figure}

In the case where the sequence of mutations all have the same mutation rate we think of the probability a $k$-step tunnel is opened by suppressing the variable $\mu$ in $g$ and recursively applying $g$ to $U(a)$. The solution can be found graphically by cobwebbing the graph of $g$ (figure \ref{fig:cobweb}) \citep{AdlerBook}.
Longer tunnels have lower probabilities, and the decrease in probability depends on the shape of $g$.  If  $g^{\prime}(0)<1$ then  probability of more  complex tunnels opening decreases towards  0 (because $g$ has no fixed point), but if $g^{\prime}(0)>1$ then the adding more intermediate mutations has a decreasing effect on the tunneling probability.
Note that $g^{\prime}(0)<1$ if $r < \frac{1}{1+\mu}$.

% So long as $U(a)$ is large enough, increasing the number of intermediate steps  decreases the probability that a tunnel is opened. 

For Wright-Fisher populations the probability of tunneling was described implicitly as a solution to a transcendental equation. However, the same recursive approach can be taken to find successive tunneling probabilities. Substituting $\omega = \mu U$ into equation (\ref{eq:ImpWF}) and noting that at a fixed point of the recursion $\alpha=1-U$ gives
\begin{equation}
\label{eq:WFrecur}
1-U = e^{-r(U)} (1-\mu U),
\end{equation}
where solving for $U$ gives the fixed point of the recursion.
Decreasing $r$ decreases the fixed point. Approximating around $U=0$ and solving gives
\begin{equation}
\label{eq;Uinf}
U_{\infty} =\frac{2(\mu + r-1)}{r(2 \mu +r)},
\end{equation}
where $U_{\infty}$ represents the fixed point of the recursion. 
If $r<1-\mu$ then there is no positive fixed point. This fixed point represents the infinite recursion for the probabilities and relies on the assumption that the time-scales of fixation and mutation can be treated separately. However,  \citet{Weissman09} found that these conditions narrow as the the length of the pathway considered increases. As the length of the pathway increases, the probability of a series of sequential fixation events increases.  However, when equation (\ref{eq;Uinf}) has no fixed point we can still infer that tunneling across long valleys will not occur. 

\section{Simulations of finite population}
The approach taken by \IMN involved first approximating the Moran process by a small time-step approximation and then using special functions and heuristic arguments to arrive at an approximation for the rate of tunneling. This method implicitly assumes large population size and explicitly ignores some higher-order terms.  My approximation explicitly assumes large population size to derive a result for the limit as population size goes to infinity. Numerical solutions can be used to evaluate the ability of these large-population size approximations to predict the rate of tunneling in finite populations. I simulated the discrete time Moran and Wright-Fisher models in order to assess the accuracy of the different approximations. My simulation draws waiting times for mutations and then tracks individuals in populations while multiple mutations are segregating. Once the secondary mutation has reached a significant size its fixation is virtually guaranteed, and the simulation is stopped. 

\begin{figure}[htbp]
\begin{center}
\includegraphics[width=1.0\textwidth]{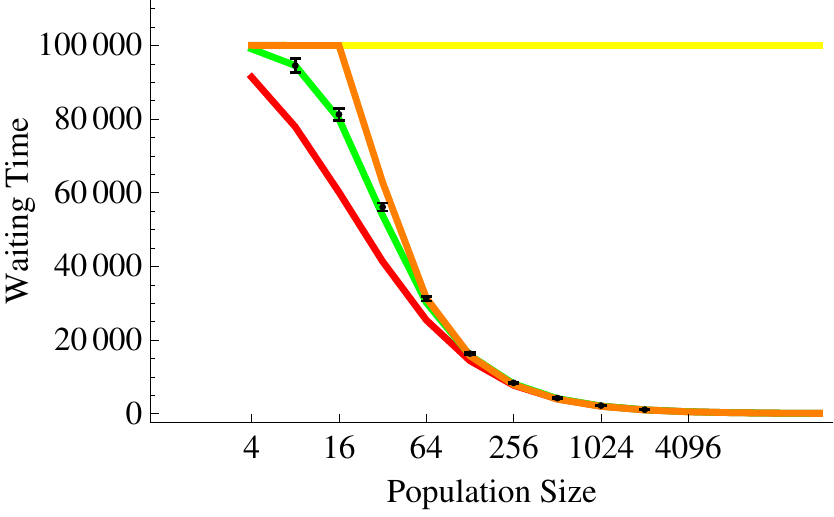}
\caption{Predicted and observed waiting times for tunneling under the Moran model. The black points show the waiting time in units of generations for a successful secondary mutation to arise along with the 95\% confidence intervals. Each parameter set was simulated $10^4$ times. The yellow curve represents the expected waiting time when the tunneling pathway is ignored. The orange curve represents the \IMN approximation (their equation 9), while the red curve represents my new approximation. The green curve was constructed by numerically solving the system for finite population size. For large population size, the two approximations agree. The parameter values were chosen to have extremely beneficial secondary mutants so that the tunneling effect is large,  $\mu_1 = 10^{-5}$, $\mu_2 = 10^{-2}$, $r = 1$, and $a = 100$. The simulations were stopped when the advantageous secondary mutation became fixed or there were more than 200 secondary mutants (the probability of loss after that point is approximately $10^{-400}$).  Similar results are obtained for $r<1$ (not shown).}
\label{fig:SimMoran}
\end{center}
\end{figure}

For the Moran model, the \IMN solution and my approximation are displaced from the numerical solution in opposite directions (figure \ref{fig:SimMoran}). 
%I used parameters that make the large-pop approximations more wrong than in the \IMN paper. I did this so you can see the results better, but you see the same thing on a smaller scale when other parameters are used. 
The \IMN approximation overestimates the waiting time to a secondary mutation. Both approximations do extremely well once population size is larger than about $100$ (for the parameter values in figure \ref{fig:SimMoran}). For smaller population sizes the variance in the waiting time is so large that, in practice, it would be hard to distinguish the alternative approximations. It is interesting to note that, in terms of displacement from the actual waiting time, the \IMN approximation performs better, even though it intentionally ignores some terms. This is apparently because the large-population approximation underestimates the waiting time (because mean fitness is not altered by the spread of mutants in the large-population limit), and by chance the terms that \IMN exclude happen to push the approximation further off.

\begin{figure}[htbp]
\begin{center}
%\begin{tabular}{cc}
\includegraphics[width=0.8\textwidth]{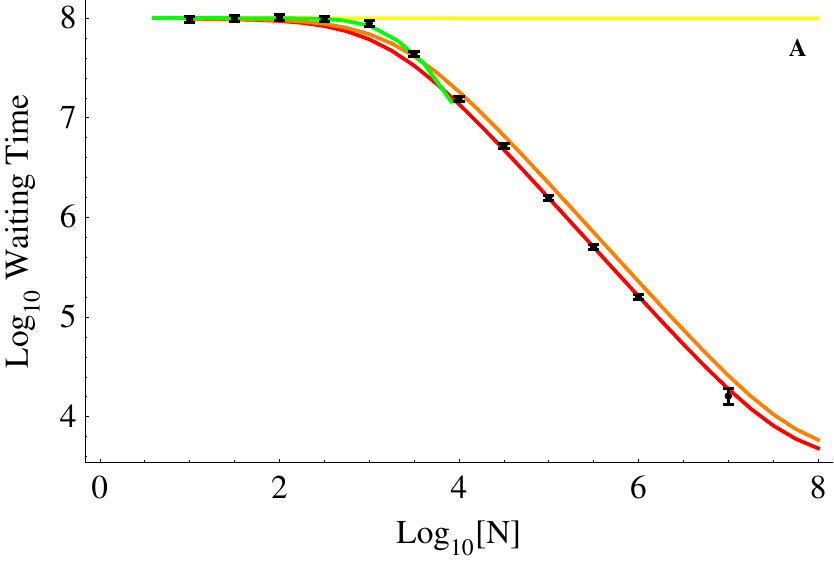} 
\includegraphics[width=0.8\textwidth]{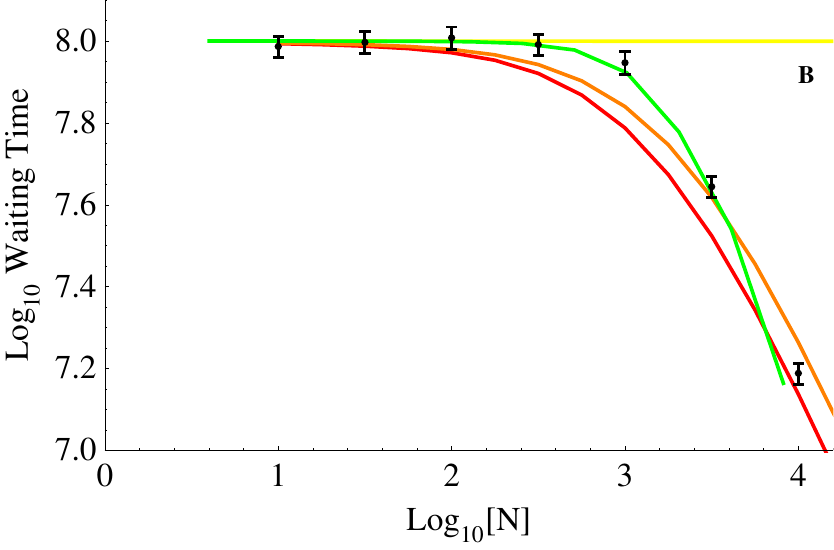}
%\end{tabular}
\caption{Predicted and observed waiting times for tunneling under the Wright-Fisher model. The black points show the waiting time for a successful secondary mutation to arise along with the 95\% confidence intervals. Because the Wright-Fisher simulations take much more time than the Moran model simulations, each parameter set was simulated only 1000 times, except for $N=10^7$ which was only simulated 100 times. The yellow curve represents the expected waiting time when the tunneling pathway is ignored. The orange curve represents the formula used by Lynch and Abegg based on the \IMN  approximation, while the red curve is based on my Wright-Fisher approximation. The green curve was constructed by numerically solving the system for finite population size. Because this requires solving a non-sparse matrix equation that is the size of the population, I was only able to numerically solve the finite population size model for population size less than $10^4$. Panel B shows the graph for these population sizes in more detail. For small population sizes, the simulated data match the numerical prediction quite well. For population sizes larger than about $10^4$ the observed values agree with my Wright-Fisher approximation. Parameter values are $\mu_1 = 10^{-8}$, $\mu_2 = 10^{-5}$, $r=1$, $a=1.01$.} %put in number of simulations
\label{fig:SimWF}
\end{center}
\end{figure}

For the Wright-Fisher model, I compared my solution with the one presented by Lynch and Abegg \cite{LynchAbegg} (figure \ref{fig:SimWF}). Lynch and Abegg applied the \IMN approximation to Wright-Fisher populations but modified the calculations to adjust for the possibility of multiple mutational hits in a single generation. My approximation for the rate of tunneling in Wright-Fisher  populations includes a factor of $\sqrt{2}$ that is left out if one simply applies the \IMN Moran approximation to Wright-Fisher populations. At small population sizes, the sequential fixation pathway dominates and the two approaches yield similar predictions. At intermediate population size the predictions are most different, but still have qualitatively similar patterns. For the parameters used in figure \ref{fig:SimWF}, the two predictions always differ by less than about $0.3\%$ and are within about 8 million generations of each other. For intermediate population sizes, the finite population matrix solution does capture the behavior of the system. As population size grows larger the simulated results move towards my approximation, just as the matrix solution converges towards my approximation.

%	: Ok, so to clarify, when r=1 the mean number of Moran muttants spawned is Np^2/2 which scales to Np/2. In the case where r<1 the mean number of mutants spawned approaches Np /(1-r) which is scaled to 1/(1-r). So the r<1 case is well explained by the mean number of mutants produced per lineage, while the r=1 case is not.

\section{Conclusions}

Multi-step evolution is becoming more widely recognized as an important component of the evolutionary process \citep{weinreich,HermissonSoft,KoppHermisson,DurrettSchmidt,LynchPNAS10,LynchAbegg}. Previous analyses  were derived for the Moran model  \citep{Iwasa04, Schweinsberg08, Weissman09} and for specific instances of the Wright-Fisher model \citep{IMN03} .  The Moran model results have been used in models of  Wright-Fisher populations \citep{LynchAbegg} and the distinction between the two models has not received much attention. The results I present here are derived using elementary methods from stochastic processes theory. I present exact expressions for the limiting case of large population size in both Moran and Wright-Fisher populations. Some methods for efficiently numerically evaluating the finite population size solutions are presented.

%Large pop size branching process approximation and WF vs Moran variance.
The Wright-Fisher and Moran models differ in two different but related ways. First, the branching process calculation of the probability of tunneling depends on a sum over the number of mutant offspring produced by a single mutant (compare equations (\ref{eq:alphaM}) and (\ref{eq:vbin})). For the same mean selection against primary mutants, the variance in  the number of offspring under the Wright-Fisher model is 1/2 as large as under the Moran model. The tunneling probability also depends on $\omega$, the product of the secondary mutation rate and the probability of fixation of the secondary mutation. Second, the fixation probability also depends on the distribution of the number of offspring and again differs by a factor of two between the models. For the same increase in fitness, the probability of fixation is twice as large under the Wright-Fisher model as under the Moran model. Because these are both introduced inside a square-root function, the total effect is that the probability of tunneling is twice as large under the Wright-Fisher model as compared to the Moran model. The general prediction is that when the offspring distribution has lower variance then the probability of tunneling will increase.

For Wright-Fisher populations, the probability of tunneling is the solution to an exponential equation which has a  simple intuitive explanation.  The total probability that no successful secondary mutants are produced is the product of the probabilities that the initial primary mutant does not immediately produce a successful secondary mutant, $1-\omega$ and the probability that none of its primary mutant progeny produce a successful secondary mutant. 

My analysis shows that tunneling in the Wright-Fisher model is more likely than in the Moran model.  Stochastic simulations show that the Wright-Fisher approximation does indeed capture the mean behavior of the evolutionary process once population size is relatively large. The improvement over applying the Moran approximation to the Wright-Fisher scenario is quite minor, but there is no added difficulty in using this correct approximation in the future.

\section*{Acknowledgements}
The work was improved by numerous discussions including those with Ricardo Azevedo, Michael Lynch, Alexey Yanchukov, Leah Johnson, Daniel Weissman, and the Theory Lunch group. The comments of three reviewers greatly improved the manuscript. This work was supported by NSF grant EF-0742582.

\appendix
\section{Approximations around small $\omega$}
\label{sec:series}
The probability of tunneling in a large Moran population is
\begin{equation}
T_{\mathrm{M}} = \sqrt{\omega(1+\omega/4)}-\frac{\omega}{2} .
\end{equation}
The second term ($-\frac{\omega}{2}$) is already linear and does not need further approximation.  We would like to approximate this function for small $\omega$ as a power series of terms $\omega^{i/2}$.  Define $f(\hat{\omega}) = (1+\hat{\omega} )$ and write the radical  as $\sqrt{\omega f(\hat{\omega})}$, where $\hat{\omega}$ will later be set equal to $\omega$. We will construct a Taylor's series approximation around $\hat{\omega}=0$. Note that $f(0)=1$ and $f'(0)=1/4$, and all higher derivatives of $f$ are 0. The series is as follows
%\begin{align}
% \sqrt{\omega f(\hat{\omega})} \approx \sqrt{\omega} +\hat{\omega} \cdot 1/2 \cdot \omega  f'(0)  (\omega f(0))^{-1/2}
%           -\frac{\hat{\omega^2}}{2!} \cdot 1/2 \cdot 1/2 \cdot  (\omega f'(0))^2 \cdot  (\omega f(0))^{-3/2} +\ldots 
% \end{align}
 \begin{align}
 \sqrt{\omega f(\hat{\omega})} \approx \sqrt{\omega} +\hat{\omega} \frac{ \omega  f'(0)  (\omega f(0))^{-1/2}}{2}
           -\frac{\hat{\omega}^2}{2!} \frac{ (\omega f'(0))^2  (\omega f(0))^{-3/2}}{2^2} +\ldots 
 \end{align}
Taking just the zero order terms gives
\begin{equation}
T_{\mathrm{M}} \approx \sqrt{\omega}-\frac{\omega}{2} +  O[\hat{\omega}] .
\end{equation}

After setting $\hat{\omega} = \omega$, the general expression for the $i$th term in the Taylor's expansion is 
\begin{equation}
s_i = \frac{(-1)^{(i+1)} \omega^{((2 i +1)/2)}  }{i!} (1/4)^i (1/2) \Pi_{j=1}^{i-1} \frac{(2 j -1)}{2} .
\end{equation}
This series converges so that  $\sum_{i=0}^{\infty} s_i = \sqrt{\omega} \sqrt{1+\omega/4}$. So a good approximation is
\begin{equation}
T_{\mathrm{M}} \approx \sqrt{\omega}\sqrt{1+\omega/4}-\frac{\omega}{2}  . %changed sign as per Weissman comment.
\end{equation}

For the Wright-Fisher model, when $r=1$,  $\alpha$ is represented implicitly by $\alpha = e^{1-\alpha}(1-\omega)$. Recall that  $T_{\mathrm{WF}}=1-\alpha$ and express it in terms of Lambert's W (the solution of $W(z) e^{W(z)} = z$) gives 
\begin{equation}
\label{eq:WFLambert}
T_{\mathrm{WF}}=1+W(\frac{\omega-1}{e}).
\end{equation}
Many methods of approximating $W$ are known, we use the power series expansion presented by \cite{CorlessW},
\begin{equation}
W(z) \approx  -1 + p - 1/3 p^2 + \cdots,
\end{equation}
where $p = \sqrt{2(e z +1)}$. Substituting this back into equation (\ref{eq:WFLambert}) gives
\begin{equation}
T_{\mathrm{WF}}\approx \sqrt{2 \omega} - \frac{2}{3} \omega + O[\omega]^{1/2} .
\end{equation}

For situations where $r<1 $ a similar approach can be used. From equation (\ref{eq:alphaM}) we have
\begin{eqnarray}
T_{\mathrm{M}} =  1- \frac{1+r(\omega+1) - \sqrt{(1-r)^2 + (2+r(\omega+2)) r \omega } }{2 r} .
\end{eqnarray}
Away from $r=1$ this can be approximated using a Taylor's series expansion to give
\begin{eqnarray}
T_{\mathrm{M}} \approx \frac{r \omega}{1-r} - \frac{(r \omega)^2}{(1-r)^3} +O[\omega]^{3} .
\end{eqnarray}

For the Wright-Fisher model with $r<1$ a regular perturbation approach can be used to the probability of a tunnel opening. First write $\alpha$ as a function of $\omega$ giving
\begin{equation}
\alpha(\omega) = e^{-r(1-\alpha(\omega))}(1-\omega),
\end{equation}
and note that $\alpha(0)=1$. Differentiating the equation with respect to $\omega$ gives
\begin{equation}
\alpha'(\omega) =  r \alpha'(\omega)  e^{-r(1-\alpha(\omega))}(1-\omega) - e^{-r(1-\alpha(\omega))}.
\end{equation}
Setting $\omega=0$ gives
\begin{align}
\alpha'(0) =  r \alpha'(\omega)  -1 \Rightarrow \alpha'(0)= - \frac{1}{1-r}.
\end{align}
Thus, 
\begin{equation}
T_{\mathrm{WF}} = 1-\alpha(\omega) \approx  \omega \frac{1}{1-r} = 2 \mu (a-1)  \frac{1}{1-r} .
\end{equation}

\section{Mean number of primary mutants spawned by a lineage destined to become extinct}
\label{sec:MeanSpawned}

\subsection{Moran populaitons}
First step analysis can be used to calculate the mean number of mutants spawned by a lineage descending from a single initial mutant. Again I condition on the eventual extinction of the mutant linage. Note that in the Moran model, time is scaled by the population size and this must be kept in mind when using results based on the number of mutants produced to estimate the probability of a secondary mutation. The first step analysis yields the system of  equations 
\begin{align}
D_1 &=  Pr(1\rightarrow 0) \frac{\pi_{0}}{\pi_1} (1+D_{0}) + Pr(1\rightarrow 1)  (1+D_{1})+ Pr(1\rightarrow 2) \frac{\pi_{2}}{\pi_1} (1+D_{2}) \\ 
D_i &=  Pr(i\rightarrow i-1) \frac{\pi_{i-1}}{\pi_i} (i+D_{i-1}) + Pr(i\rightarrow i) (i+D_{i})+ Pr(i\rightarrow i+1) \frac{\pi_{i+1}}{\pi_i} (i+D_{i+1}) \\
D_{N-1} &=  Pr( N-1 \rightarrow N-2) \frac{\pi_{N-2}}{\pi_{N-1}} (N-1+D_{N-2}) + Pr(N-1\rightarrow N-1)  (N-1+D_{N-1}) ,
\end{align}
where $D_i$ is defined as the expected cumulative weight of the number of descendants from a mutant lineage starting with $i$ mutants, conditioned on eventual extinction of the lineage \citep{Weissman09}. This weight represents the mutational opportunity for the lineage of primary mutants and is scaled by the population size.   The boundary conditions are that $D_0 = 0$ while $D_N$ does not need to be defined since no transition to state $N$ is possible.
In the neutral case where $r=1$ this reduces to 
\begin{align}
\label{eq:NeutMoranD}
D_1 &=  \frac{1}{N} (1+D_{0}) + \left( 1- \frac{2(N-1)}{N^2}\right)  (1+D_{1})+  \frac{N-2}{N^2}    (1+D_{2}) \\ 
D_i &=   \frac{i (N-i+1)}{N^2} (i+D_{i-1}) + \left(1-\frac{2 i (N-i)}{N^2} \right)(i+D_{i})+ \frac{i (N-i-1)}{N^2} (i+D_{i+1}) \nonumber \\
D_{N-1} &=  \frac{2 (N-1)}{N^2}  (N-1+D_{N-2}) + \left(1-\frac{2  (N-1)}{N^2} \right)  (N-1+D_{N-1}) \nonumber .
\end{align}
The solution $D_i = i N^2 /2$ satisfies system (\ref{eq:NeutMoranD}).  Because generations are scaled in terms of $N$ time steps in the Moran model we can say that on average a single mutant produces an effective number of $N/2$ descendants before going extinct. That is, the sum of the time that primary mutant descendants are alive is $N/2$ generations.

In the case where $r<1$ we can write the system for finite population size but I have found no simple way of expressing its solution. In the limit of large population size we can make use of the fact that a lineage starting with $i$ mutants must produce $i$ times as many descendants as a lineage starting with 1 mutant. I define $D^*_i $ as the number of descendants produced scaled to the population size such that $D_i = D^*_i  N$. Solving for $D^*_{i+1}$ and taking the limit as $N \rightarrow \infty $ gives
\begin{align*}
D^*_{i+1} &=  \frac{(r+1) D^*_{i} - D^*_{i-1} - 1}{r} .
\end{align*}
The additional conditions that $\frac{D^*_i}{i} =\frac{D^*_{i-1}}{i-1} =\frac{D^*_{i+1}}{i+1}$ implies that
\begin{equation}
D^*_{i}=\frac{i}{1-r}.
\end{equation}
Thus, for the Moran model, the scaled number of mutants produced by a single mutant approaches $1/(1-r)$. For finite populations, the value is within 1\% of this limit for populations larger than 100 when $r<0.75$.

%Note that Appendix calcs has these

\subsection{Wright-Fisher populations}
For Wright-Fisher populations I calculate the number of descendants when the primary mutation is neutral. Consider a population with
$N$ haploid genotypes. The first step equations are
\begin{align*}
D_i &= \sum_{j=0}^{N}  Pr(i\rightarrow j) \frac{\pi_{j}}{\pi_i} (i+D_{j}),\\
Pr(i\rightarrow j) &= \binom{ N}{j} \left(\frac{i}{ N}\right)^j  \left(\frac{ N-i}{ N}\right)^{ N-j} ,\\
\pi_{i} &= \frac{ N-i}{ N} .
\end{align*}
Because the mutant allele is neutral there is no effect on mean fitness as the number of primary mutants changes in the population. This means that, conditioned on non-fixation of the primary mutation, the number of descendants left by $i$ primary mutants must just be $i$ times the number left by 1 primary mutant. Thus $D_i = D_1 i$. Inserting this relationship back into the system of equations gives
\begin{equation*}
D_i= i D_1   = \frac{1}{N-i} \sum_{j=0}^{N} \binom{ N}{j} \left(\frac{i}{ N}\right)^j  \left(\frac{ N-i}{ N}\right)^{ N-j} (N-j) (i+j D_1),
\end{equation*}
which is can be written as the sum of  terms involving the first and second (non-central) moments of the binomial distribution with parameters $N$ and $i/N$. Simplification gives 
\begin{equation*}
 D_1 i  = i \left(1+ \frac{D_1(N-1)}{N} \right).
\end{equation*}

Solving for $D_1$ gives $D_1 = N$. Thus 
\begin{equation}
D_i = N i.
\end{equation}
So for a haploid Wright-Fisher population the mean number of neutral mutants spawned by a lineage destined to extinction is equal to the number of haploid genomes present in the population.

For deleterious mutations it is not possible to write the conditional process for finite populations because no closed-from solution for the fixation probabilities is available. Instead, I calculate the average number of descendants in a large population using the Poisson approximation. The first step equations are
\begin{align*}
D_i &= \sum_{j=0}^{N}  Pr(i\rightarrow j)  (i+D_{j}),\\
Pr(i\rightarrow j) &= \frac{e^{- i r} (i r)^j}{j!}.
\end{align*}
In very large populations the branching approximation applies and  $D_i = i D_1$,  yielding
\begin{equation*}
D_i  = D_1  i = i + D_1  i r.
\end{equation*}
Solving for $D_1$ shows that 
\begin{equation}
D_i  = \frac{i}{1-r}.
\end{equation}

\begin{center}
	\begin{tabular}{ |c | l |}
		\hline
		$\mu_1$ 	&probability that a wild-type allele mutates to produce a primary mutation \\ \hline
		$\mu_2$	&probability that a primary mutant allele mutates to produce a secondary mutation \\ \hline
		$N$	& Number of haploid genomes in the population  \\ \hline
		$U(x)$	& probability an allele with relative fitness $x$ becomes fixed when initially present as a single copy \\ \hline
		$U_{\infty}$	& probability that a tunnel of infinitely many steps will open. \\ \hline

		$r$		&  fitness of primary mutants relative to the wild-type \\ \hline
		$a$		&  fitness of secondary mutants relative to the wild-type \\ \hline
		$S_1$	& probability that a primary mutation destined to become fixed arises in a given generation \\ \hline
		$S_2$	& probability that a secondary mutation destined to become fixed arises in a population \\
				& composed entirely of primary mutants \\ \hline
		$T$		& probability, in a population of wild-type alleles, of a primary mutant destined to (before \\
				& the primary mutant becomes fixed) give rise to a secondary mutant that then becomes fixed \\ \hline 
		$\pi_i$	& probability of eventual extinction of a lineage descending from $i$ primary mutants \\ \hline
		$\omega$	& composite parameter equal to $\mu_2 U(a) $ \\ \hline
		$v_i $	&  probability that no successful secondary mutations are produced from a lineage descending \\
				& from $i$ primary mutants, conditional on the non-fixation of the primary mutation \\ \hline
		$\vect{v}$	& vector of the probability that no successful secondary mutations are produced \\ \hline
		$\tilde{v}_i$& unconditional probability that no successful secondary mutations are produced from a lineage \\
				& descending from $i$ primary mutants\\ \hline 
		$\alpha$ & probability that no successful secondary mutants are spawned from a lineage descending from \\
				& a single primary mutant \\ \hline
		$\alpha_{\mathrm{IMN}}$ & approximate $\alpha$ derived by \IMN  \\ \hline
		$T_{\mathrm{M}}$ 
		                  & for the Moran model, the probability that a single primary mutant will produce a lineage\\
		                  & that produces a successful secondary mutant. \\ \hline 
		$T_{\mathrm{WF}}$ 
		                  & for the Wright-Fisher model, the probability that a single primary mutant will produce a lineage\\
		                  & that produces a successful secondary mutant. \\ \hline 
	\end{tabular}
\end{center}

\clearpage		
\section*{Figures}
\clearpage

\clearpage

\renewcommand\refname{LITERATURE CITED} 

\bibliography{Tunnelling}

\end{document}